\documentclass{svjour3}                     
\smartqed  
\usepackage{graphicx}
\begin{document}

\title{Proposed New Antiproton Experiments at Fermilab \thanks{To appear in Proceedings of IXth International Conference on Low Energy Antiproton Physics (LEAP'08), Vienna, Austria, September 16 to 19, 2008.}
}


\author{Daniel M. Kaplan 
}


\institute{D. M. Kaplan \at
              Illinois Institute of Technology \\
              Tel.: +1 312-567-3389\\
              Fax: +1 312-567-3289\\
              \email{kaplan@iit.edu}           
}

\date{Received: date / Accepted: date}

\maketitle

\begin{abstract}
Fermilab operates the world's most intense source of antiprotons. Recently various 
experiments have been proposed that can use those antiprotons either parasitically during 
Tevatron Collider running or after the Tevatron Collider finishes in about 2010. We discuss 
the physics goals and prospects of the proposed experiments. 

\keywords{Antiproton \and Antihydrogen \and Antimatter \and Charm \and Charmonium \and CPT \and Gravity \and Hyperons}
\end{abstract}

\section{Introduction}
\label{intro}
The world's highest-energy and highest-intensity antiproton source is at Fermilab (see Table~\ref{tab:sens-comp}). Having previously supported medium-energy antiproton fixed-target experiments (including the charmonium experiments E760 and E835), it is now 100\% dedicated to providing luminosity for the Tevatron Collider. At CERN, the LEAR antiproton storage ring was decommissioned in 1996;\footnote{LEAR was turned off in spite of its review committee's recommendation  that it be allowed to complete its planned program of research; the rationale was to free up expert manpower for LHC work. The ``ground rules" for the AD design accordingly required operability by as small a crew as possible.}
its successor facility, the Antiproton Decelerator (AD), provides antiproton beams at  momenta of 100 and 300\,MeV/$c$, at intensities up to $\approx2\times10^7$ per minute~\cite{Eriksson}. 
It is noteworthy that Germany has embarked on  a $\approx$billion-Euro upgrade plan for the GSI-Darmstadt nuclear-physics laboratory that includes construction of 30 and 90\,GeV rapid-cycling synchrotrons and low- and medium-energy antiproton storage rings~\cite{FAIR}. These facilities address an interesting and varied research program in nuclear and particle physics and beyond. 
\begin{table}\begin{center}
\caption{Antiproton energies and intensities at existing and future facilities.}\label{tab:sens-comp}\vspace{0.1in}
\begin{tabular}{lccccc}
\hline\hline & ${\overline{p}}$ K.E. & \multicolumn{2}{c}{{Stacking:}} &{Hours} & ${\overline{p}/}${Yr}\\
\raisebox{1.5ex}[0pt]{{Facility}} &  (GeV)& {Rate} ${(10^{10}/}${hr)}  & { Duty Factor} & {/Yr} & ${(10^{13})}$ 
\\
\hline\hline
CERN AD & 0.005, 0.047 & -- & -- & 3800 &0.4 \\
FNAL  (Accumulator) &$\approx$\,3.5--8& 20 & 15\% & 5550 &17 \\
FNAL  (New Ring) &2--20? & 20 & 90\%& 5550 & 100 \\
FAIR ($\stackrel{>}{_\sim}\,$2015) & 2--15 &  3.5 & 90\% & 2780$^*$ & 9\\
\hline\hline
\end{tabular}
\end{center}
~~$^*$ The lower number of operating hours at FAIR compared with that at other facilities arises from medium-energy antiproton operation having to share time with other programs.
\vspace{-.25in}
\end{table}

\section{Physics Overview}
A number of intriguing recent discoveries  can be elucidated at such a facility, foremost among which is charm mixing~\cite{HFAG}. The key question is whether there is  new physics in charm mixing; the signature for this is {\em CP} violation~\cite{Bigi-Uraltsev-Petrov}. The search for new physics in $B$ and $K$ mixing and decay has so far come up empty. Thus it behooves us to look elsewhere as well. As pointed out by many authors, charm is an excellent venue for such investigation: It is the only up-type quark for which such effects are possible, and standard-model backgrounds to new physics in charm are suppressed by small CKM-matrix elements and the fact that the $b$ quark is the most massive one participating in loop diagrams~\cite{Buchalla-etal}. We argue below that a charm experiment at the Fermilab Antiproton Source might be the world's most sensitive.

Other topics of interest include such states as the 
$X(3872)$
in the charmonium region~\cite{ELQ}, observed by several groups, as well as the investigation of possible new-physics signals observed in the HyperCP experiment at Fermilab: evidence for {\em CP} violation~\cite{BEACH08} and flavor-changing neutral currents~\cite{Park-etal} in hyperon decay. In addition, the $h_c$ mass and width, $\chi_c$ radiative-decay angular distributions, and ${\eta^\prime_c}(2S)$ full and radiative widths, important parameters of the charmonium system that remain to be precisely determined~\cite{QWG-Yellow}, are well suited to the 
${\overline p}p$ technique~\cite{E835}. 

Charm particles can be pair-produced in $\overline{p}p$ or $\overline{p}N$ collisions at and above the $\psi(3770)$ resonance. There is an enormous cross-section advantage relative to $e^+e^-$ colliders: charm hadroproduction cross sections are typically ${\cal O}(10\,\mu$b), while $e^+e^-$ cross sections are ${\cal O}$(1\,nb). Against this must be weighed the  $e^+e^-$ luminosity advantage, typically ${\cal O}(10^2)$, and the lower background rates in $e^+e^-$ experiments. Charm hadroproduction at high energies comes with the advantage of longer decay distances, but the countervailing disadvantage of higher multiplicity ($\langle n_{ch}\rangle\sim10$) in the underlying event. We expect that the low charged-particle multiplicity  ($\langle n_{ch}\rangle\approx2$)  in $\overline{p}p$ collisions  somewhat above open-charm threshold will enable charm samples with cleanliness comparable to that at the $B$ factories, with the application of only modest cuts, and hence, high efficiency. The competition for this program is a possible ``super-$B$ factory." 

By scanning the Antiproton Accumulator beam energy across the resonance, Fermilab experiments E760 and E835 made the world's most precise measurements of charmonium masses and widths~\cite{E835}. 
Besides this precision, the other key advantage of the antiproton-annihilation technique is its ability to produce charmonium states of all quantum numbers, in contrast to $e^+e^-$ machines which produce primarily $1^{--}$ states and the few states that couple directly to them, or (with relatively low statistics) states accessible in $B$ decay or in $2\gamma$ production.

The E835 apparatus did not include a magnet, thus various cross sections needed to assess the performance and reach of a new experiment remain unmeasured. However, they can be estimated with some degree of confidence. We are proposing to assemble, quickly and at modest cost, an ``upgraded E835" spectrometer that includes a magnet. If these cross sections are of the expected magnitude, it should be possible with this apparatus to make the world's best measurements of charm mixing and {\em CP} violation, as well as of the other effects mentioned above. (To take full advantage of the capabilities of the Fermilab Antiproton Source, a follow-on experiment in a new, dedicated ring \`a la Table~\ref{tab:sens-comp} might  then be designed for even greater sensitivity.)

\section{Proposed antiproton experiments at Fermilab}

\subsection{Medium-energy $
{\overline p}
 p$-annihilation experiment}

\begin{figure}
\vspace{-.05in}
\centerline{\hspace{-.25in}
\includegraphics[width=.66\linewidth]{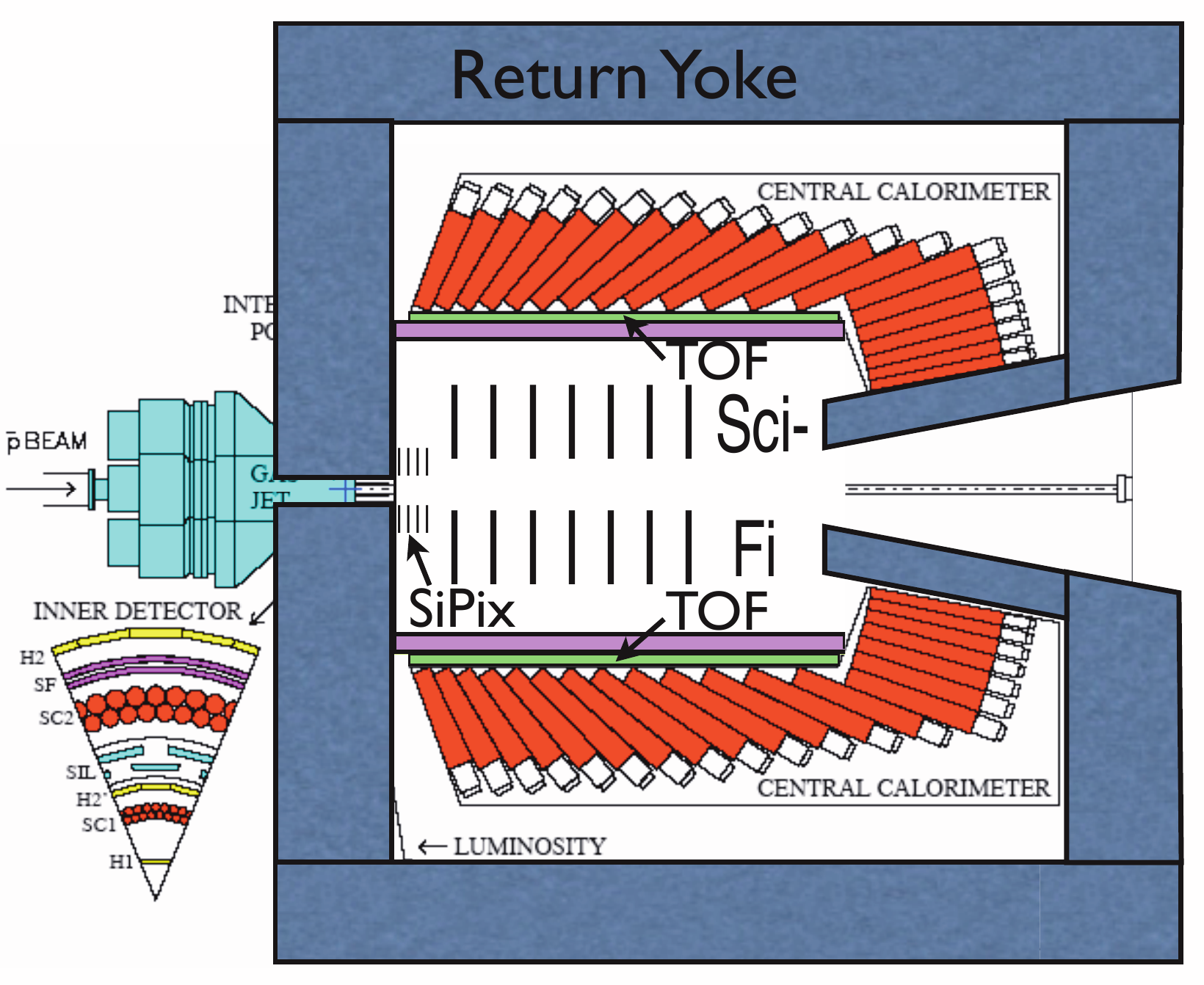}
}
\caption{Sketch of upgraded E835 apparatus as discussed in text: 1\,T solenoid shown in magenta, TOF counters in green. Return yoke (needed to assure operability of calorimeter phototubes) should consist of as little iron as necessary.}\label{fig:E835-upgrade}
\end{figure}

By adding a small magnet, tracking and vertex detectors, and TOF counters to the E835 calorimeter as in Fig.~\ref{fig:E835-upgrade}, plus modern, high-bandwidth triggering and data-acquisition systems, several important topics can be studied. We assume $\overline p$$p$ or $\overline p$$N$ luminosity of $2\times10^{32}$\,cm$^{-2}$s$^{-1}$, one order of magnitude beyond that of E835, which can be  accomplished by use of a denser internal target than the E835 hydrogen cluster jet~\cite{E835}.

\paragraph{Charm mixing and  CP violation}

After a more than 20-year search, $D^0$--${\overline D}{}^0$ mixing is now established  at 9.2 standard deviations~\cite{HFAG}, thanks mainly  to the $B$ factories. The level of mixing is consistent with the wide range of standard-model predictions~\cite{Bigi-Uraltsev-Petrov}; however, this does not preclude a significant and potentially detectable contribution from new physics~\cite{Grossman-etal}. Since some new-physics models predict different effects in the charge-2/3 (``up-type") quark sector than in the down-type~\cite{Grossman-etal}, it is important to carry out such  studies of charm mesons\,---\,the only up-type system for which  meson mixing can occur.

The $\overline{p}p$ annihilation cross section to open charm could be  substantial; for example, a recent estimate gives $\sigma(\overline{p}p\to D^{*0}{\overline D}{}^0)\approx1.3\,\mu$b at $\sqrt{s}=4.2$\,GeV~\cite{Braaten}. At ${\cal L}=2\times10^{32}$\,cm$^{-2}$s$^{-1}$, this represents some $5\times10^9$ events per year, substantially exceeding each year the integrated sample ($10^9$ events) 
available at the $B$ factories. Since there will  also be $D^{*\pm}D^\mp$, $D^*\overline{D}{}^*$, and $D\overline{D}$ events,  the total charm sample will be even larger; with the use of a target nucleus heavier than hydrogen, the charm-production $A$-dependence~\cite{A-dep} could enhance statistics by an additional factor of a few. Such a target could also localize primary interactions to an ${\cal O}(\mu$m)-sized region, allowing the $D$-meson decay distance to be cleanly resolved. Medium-energy ($p_{\overline{p}}\approx8\,$GeV/$c$) $\overline{p}N$ annihilation may thus be the optimal way to study charm mixing and search for possible new-physics contributions via the clean signature~\cite{Petrov} of charm {\em CP} violation (CPV).

\paragraph{Hyperon CP violation and rare decays}

\begin{figure}
\centerline{
\includegraphics[width=0.44\linewidth]{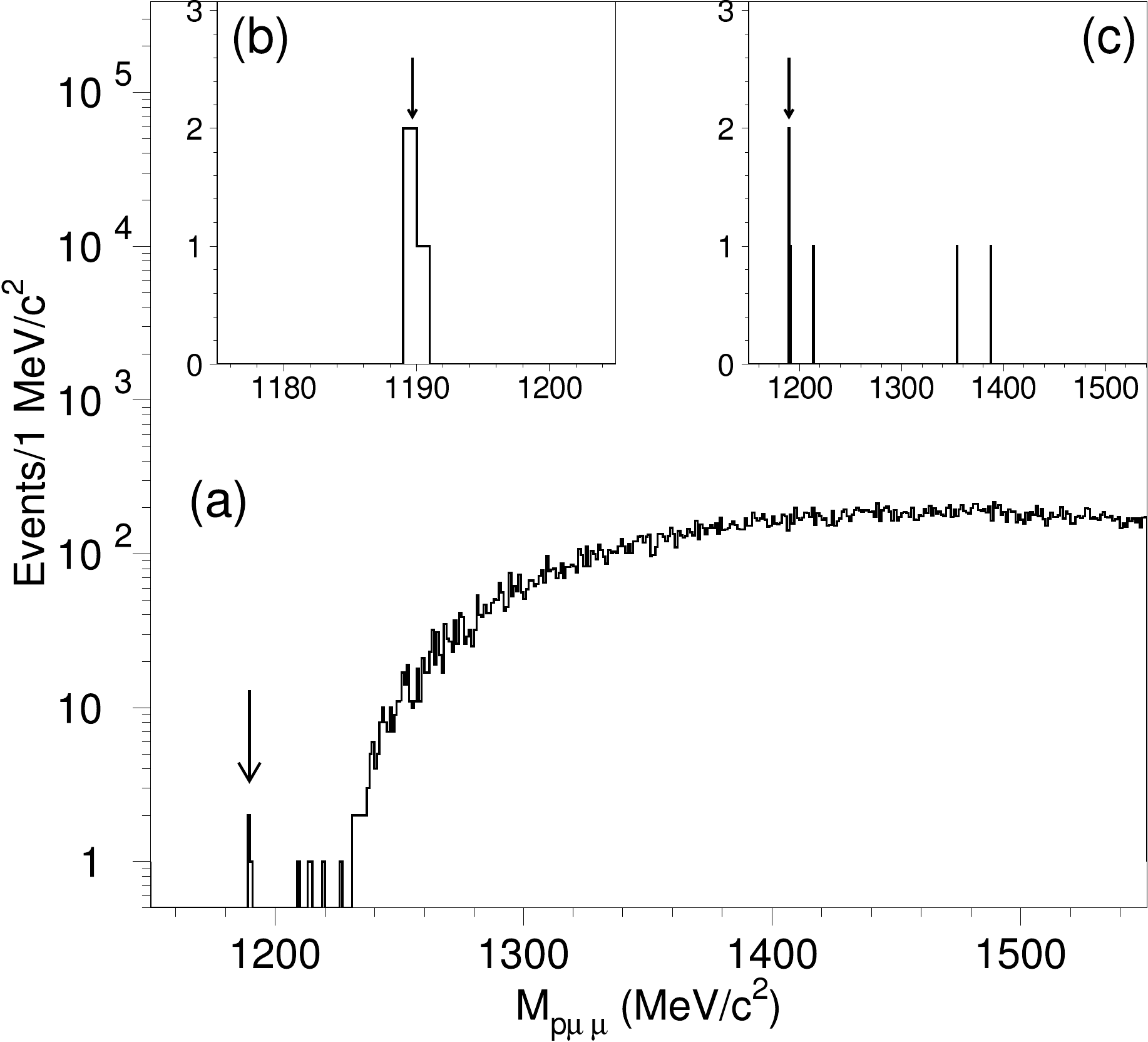}~~\includegraphics[width=0.55\linewidth]{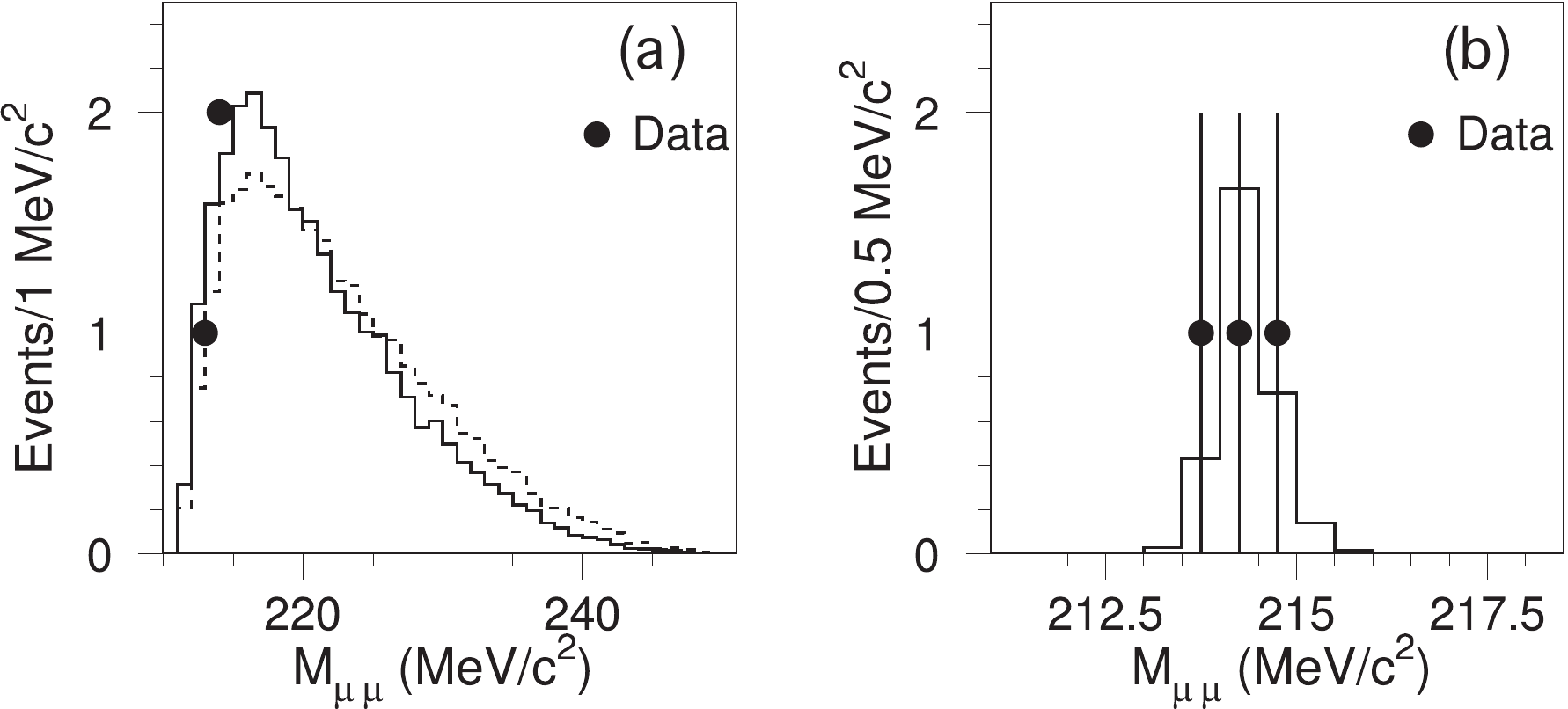}}
\caption{(Left) Mass spectrum for 3-track final states consistent with being single-vertex $p\mu^+\mu^-$  events in HyperCP positive-beam data sample: (a) wide mass range (semilog scale); (b) narrow range around $\Sigma^+$ mass; (c) after application of additional cuts as described in Ref.~\protect\cite{Park-etal}. (Arrows indicate mass of $\Sigma^+$.) (Right) Dimuon mass spectrum of the three HyperCP $\Sigma^+\to p\mu^+\mu^-$ candidate events compared with Monte Carlo spectrum assuming (a) SM virtual-photon form factor (solid) or isotropic decay (dashed), or (b) decay via a narrow resonance $X^0$.
\label{fig:Sigpmumu}}
\end{figure}

The Fermilab HyperCP Experiment~\cite{E871} amassed the world's largest samples of hyperon decays, including $2.5\times10^9$ reconstructed ${}^{^(}\overline{\Xi}{}^{^{\,)\!}}{}^\mp$ decays and $10^{10}$ produced $\Sigma^+$. HyperCP observed unexpected signals at the $\stackrel{>}{_\sim}$\,2$\sigma$ level  for possible new physics in the rare hyperon decay $\Sigma^+\to p\mu^+\mu^-$~\cite{Park-etal} and the  ${}^{^(}\overline{\Xi}{}^{^{\,)\!}}{}^\mp\to{}^{^(}\overline{\Lambda}{}^{^{\,)}}\pi^\mp\to 
{}^{^{(}}\overline{p}{}^{^{\,)}} 
\pi^\mp\pi^\mp$ {\em CP} asymmetry~\cite{BEACH08}. Since the $\overline{p}p\to\Omega^-{\overline\Omega}{}^+$ threshold lies in the same region as the open-charm threshold, the proposed  experiment  can further test these observations using ${}^{^(}\overline{\Omega}{}^{^{\,)\!}}{}^\mp\to{}^{^(}\overline{\Xi}{}^{^{\,)\!}}{}^\mp\mu^+\mu^-$ decay and potential ${}^{^(}\overline{\Omega}{}^{^{\,)\!}}{}^\mp$ CPV~\cite{OmegaCP}. While the $\overline p$$p\to \Omega^- {\overline \Omega}{}^+$ cross section has not been measured, by extrapolation from $\overline p$$p\to \Lambda \overline \Lambda$ and $\overline p$$p\to \Xi^- {\overline \Xi}{}^+$ one obtains an estimate just above threshold of $\approx$\,60\,nb, implying $\sim10^8$ events produced per year. In addition the measured $\approx 1\,$mb cross section for associated production of inclusive hyperons~\cite{Chien-etal} would mean $\sim10^{12}$ events produced per year, which could  directly confront the HyperCP evidence (at $\approx$\,2.4$\sigma$ significance) for a possible new particle of mass 214.3\,MeV/$c^2$ in the three observed $\Sigma^+\to p\mu^+\mu^-$ events (Fig.~\ref{fig:Sigpmumu}).\footnote{Such a particle, if confirmed, could be evidence for nonminimal SUSY~\protect\cite{Gorbunov}.}   Further in the future, the dedicated $\overline{p}$ storage ring of Table~\ref{tab:sens-comp} could decelerate antiprotons to the $\Lambda\overline{\Lambda}$, $\Sigma^+\overline{\Sigma}{}^-$, and $\Xi^-\overline{\Xi}{}^+$ thresholds, where an experiment at $10^{33}$ luminosity could amass the clean, $>10^{10}$-event samples needed to confirm or refute the HyperCP evidence for {\em CP} asymmetry in $\Xi^\pm$ decay~\cite{BEACH08}.

\paragraph{Precision charmonium measurements}

Using the Fermilab Antiproton Source,  experiments E760 and E835 made the world's most precise measurements of charmonium masses and widths~\cite{E835}. 
Although charmonium has by now been extensively studied, a number of questions remain, including the nature of the mysterious $X(3872)$ state~\cite{ELQ} and improved measurement  of $h_c$ and $\eta^\prime_c$ parameters~\cite{QWG-Yellow}. The unique precision of the ${\overline p}p$ energy-scan technique is ideally suited to making the precise mass and width measurements needed to test the intriguing hypothesis that the $X(3872)$ is a $D^{*0}\overline{D}{}^0$ molecule~\cite{molecule}.

\subsection{Antihydrogen experiments} 
\paragraph{Antihydrogen-in-flight CPT tests}

The study of antihydrogen atoms in flight may be a way around some of the difficulties encountered in the CERN trapping experiments. First steps in this direction were taken by PS210 at LEAR~\cite{Baur} and Fermilab E862~\cite{Blanford}, which observed formation of antihydrogen in flight in the mid-1990s. Methods to measure the antihydrogen Lamb shift and fine structure (the $2s_{1/2}$--$2p_{1/2}$ and $2p_{1/2}$--$2p_{3/2}$ energy differences) were subsequently worked out~\cite{Blanford-Lamb-shift}. Progress toward this goal may be compatible with normal Tevatron Collider operations\,---\,a possibility  currently under investigation. If the feasibility of the approach is borne out by further work, the program could continue into the post-Tevatron era. Sensitivity at the parts-per-billion level ($\sim$\,10$^{-9}$ of the $2S$ energy) may be possible\,---\,not the $\sim10^{-14}$ envisioned for the AD program~\cite{AEGIS} but a valuable first step.

\paragraph{Antimatter gravity experiment}

While General Relativity predicts that the gravitational forces on matter and antimatter should be identical, no direct experimental test of this prediction has yet been made~\cite{Fischler-etal}. Attempts at a quantized theory of gravity generally introduce non-tensor forces, which could cancel for matter-matter and antimatter-antimatter interactions but add for matter-antimatter ones. In addition, possible ``fifth forces" or non-$1/r^2$ dependence have been discussed. Such effects can be sensitively sought by measuring the gravitational acceleration of antimatter in the field of the earth. While various such experiments have been discussed for many years, one\,---\,measurement of the gravitational acceleration of antihydrogen\,---\,has only recently become feasible and is now proposed both at CERN and at Fermilab~\cite{AEGIS,LoI}.

The principle of the Antimatter Gravity Experiment (AGE) is to form a beam of slow ($\approx1\,$km/s)  antihydrogen in a Penning trap and pass the beam through a $\approx1$-m-long Mach-Zehnder interferometer. The phase of the interference pattern can be measured to a small fraction of the ($\approx1\,\mu$m) grating period, and measurement of the phase vs.\ the speed of the atom determines $\bar g$. Simulation shows that $\bar g$ can be measured to 0.6\% of $g$ with one-million antihydrogen ($\overline H$) atoms incident on the interferometer; this could be done parasitically during the Tevatron run. The proposed AGE goal is a $10^{-4}$ measurement, requiring $10^{10}$ $\overline H$ atoms. Given the expected $\stackrel{>}{_\sim}$\,10$^{-5}$ antiproton-trapping and $\overline H$-formation efficiency, this can be accomplished at the Antiproton Source in a few-month dedicated run.

\section{Outlook}

With the end of the Tevatron Collider program in sight, new and unique measurements are possible at the Fermilab Antiproton Source~\cite{New-pbar}. Such a program can substantially
 broaden the clientele and appeal of US particle physics.

%
%

\begin{acknowledgements}
The author thanks his pbar collaborators,\footnote{Arizona: A.~Cronin; Riverside: A.~P.~Mills, Jr.; Cassino: G.~M.~Piacentino; Duke: T.~J.~Phillips; FNAL: G.~Apollinari, D.~R.~Broemmelsiek, B.~C.~Brown, C.~N.~Brown, D.~C.~Christian, P.~Derwent, M.~Fischler, K.~Gollwitzer, A.~Hahn, V.~Papadimitriou, R.~Stefanski, J.~Volk, S.~Werkema, H.~B.~White, G.~P.~Yeh; Ferrara: W.~Baldini, G.~Stancari, M.~Stancari; Hbar Tech.: J.~R.~Babcock, S.~D.~Howe, G.~P.~Jackson, J.~M.~Zlotnicki; IIT: D.~M.~Kaplan, T.~J.~Roberts, H.~A.~Rubin, Y.~Torun, C.~G.~White; KSU: G.~A.~Horton-Smith, B. Ratra; KyungPook: H.~K.~Park; Luther Coll.: T.~K.~Pedlar; Michigan: H.~R.~Gustafson, M.~Longo, D.~Rajaram; Northwestern: J.~Rosen; Notre Dame: M. Wayne; SMU: T.~Coan; SXU: A. Chakravorty; Virginia: E.~C.~Dukes; Wayne State: G.~Bonvicini.
}
and E. Braaten, E. Eichten, and C. Quigg for useful conversations. Work supported by Department of Energy grant DE-FG02-94ER40840.
\end{acknowledgements}



\end{document}